\documentclass[a4paper,11pt]{article}
\usepackage[colorlinks=true, allcolors=teal]{hyperref} 
\usepackage[utf8]{inputenc} 	
\usepackage{amsmath} 			
\usepackage{graphicx} 			
\usepackage[UKenglish]{babel} 	
\usepackage[font=small,labelfont=bf, margin=1cm]{caption} 			
\usepackage{url} 				
\usepackage{listings}            
\usepackage{color}				
\usepackage{authblk}            
\usepackage[usenames,dvipsnames,svgnames,table]{xcolor} 
\counterwithin{figure}{section}
\usepackage{booktabs}

\def\0{$0\nu\beta\beta$ }   

\usepackage[a4paper,top=2cm,bottom=2cm,left=3cm,right=3cm,marginparwidth=1.5cm]{geometry}
\usepackage{lastpage} 
\usepackage{fancyhdr}
\usepackage{tocloft}

\title{NEXT Simulation Dataset for AI Summer School UC Irvine 2026}
\author[1]{Krishan Mistry (on behalf of the NEXT Collaboration)\thanks{krishan.mistry@uta.edu} }
\affil[1]{Department of Physics, University of Texas at Arlington, Arlington, TX 76019, USA}

\begin{document}

\maketitle

\begin{abstract}

This document details the dataset release of simulated $0\nu\beta\beta$ and background events originating from the decay of $^{214}$Bi in high-pressure xenon gas, describing events similar to those produced in the NEXT detector. This release is part of the Neutrinoless Double Beta Decay ($0\nu\beta\beta$) AI Summer School held on June 20-21 2026 at the University of California, Irvine. 

\end{abstract}

\section{The NEXT Experiment}

The NEXT experiment is a high-pressure xenon gaseous time projection chamber (TPC) detector located in the Canfranc Underground Laboratory searching for neutrinoless double beta decay with the $^{136}$Xe isotope. 

A brief description of the detector operation is as follows. Charged particles traversing the xenon gas ionize and excite the xenon. Excitations lead the production of prompt scintillation photons (S1) which are detected by photomultiplier tubes (PMTs) which located at the detector sides. The ionization electrons are drifted to an anode plane where they are amplified in a region of high electric field into scintillation photons with proportional gain (S2). Silicon Photomultipliers (SiPMs) located at the anode plane use the S2 light to produce detailed 2D images of the tracks while the PMTs record the energy of the charged particles with an energy resolution of less than 1\% full width half-maximum (FWHM) at 2.5 MeV. The combination of the 2D information from the SiPMs and timing difference between S1 and S2 light signals allow the event to be reconstructed in 3D. 

\section{Signal and Background}
When a neutrinoless double beta decay event occurs, two electrons are produced with a summed energy of 2.458 MeV. As these electrons scatter and loose energy in the gas they produce a tortuous topology while depositing energy and slowing down. As they come to a stop, the angle of scatters increase where they form a dense region of ionization charge near the ends of the track. This is known as a ``blob", and is one of the key signatures in the NEXT detector for identifying stopping electrons. 

One of the major backgrounds in the NEXT experiment originates from the decay of $^{214}$Bi which is found in the natural radioactive decay chain of $^{238}$U. When $^{214}$Bi decays to $^{214}$Po, one deexitation gamma ray is released with a branching ratio of 1.5\% at an energy of 2.447 MeV, right near the Q-value of neutrinoless double beta decay of $^{136}$Xe. This gamma ray can do multiple Compton scatters and/or photoelectric effect depositing all of its energy in the detector becoming a significant background. All these interaction mechanisms either lead to the production of multiple energy deposits separated in space and/or the production of a single electron near the Q-value. 

The NEXT experiment uses the topology to identify single-tracks and single-blob events to discriminate signal (2e$^-$) vs background (1e$^-$) events. This is a perfect use-case for machine learning (ML) training whereby the task is to simply identify whether there are one or two electrons in the event. 

Sometimes the MeV-scale electrons can also radiate bremsstrahlung leading to multi-site events in the case of signal or hard-scatter producing high-energy delta rays that can produce a blob leading to two-electron signatures in the backgrounds. This physics adds a little more complexity to the clear identification of signal vs background; however, there are still markers to identify such events such as looking for forks in the trajectories or looking at variable such as the track length. 

\section{Monte Carlo Simulation}

A Monte Carlo simulation has been created for creating a ML model. The simulation includes 648,793 signal neutrinoless double beta decay events occurring directly in the xenon gas and 516,696 background events modelling $^{214}$Bi decays occurring in the walls of the detector. The simulation is done using NEXUS ~\cite{nexus}, a Geant4~\cite{Agostinelli:2002hh} based simulation. 

The geometry used is a cylinder with a diameter and length of 3.6 m at 5 bar pressure. The cylinder is surrounded by 4 cm of copper to model the inner shielding. The geometry size is typical for a detector accommodating a 1-tonne source mass. Signal events are generated using the DECAY0 generator in NEXUS in the gas volume, while background events are generated in the 4 cm copper shielding, accounting for the full $^{214}$Bi decay spectrum. 

\begin{figure}[hbt]
\centering
\includegraphics[width=\textwidth]{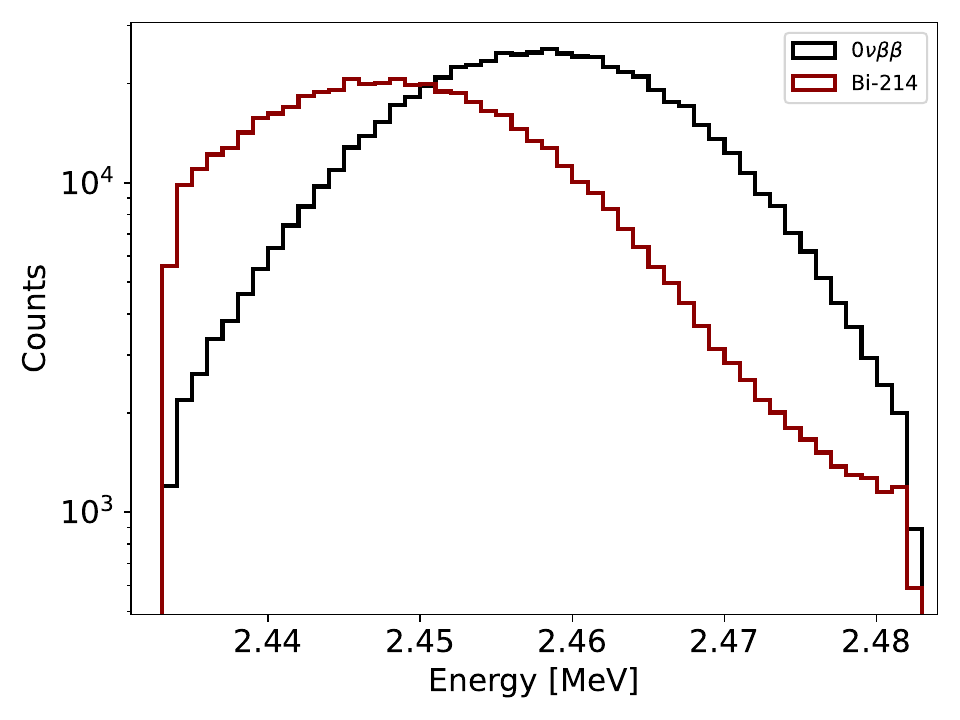}
\caption{\label{fig:energy} The energy spectrum of signal and background events in this simulation. Each entry in this histogram is the summed energy of all hits across each unique event.}
\end{figure}

The event hits are smeared to account for an energy resolution of 1\% FWHM. Only events depositing energy in a window between 2.433 - 2.483 MeV are saved. This corresponds to a 1\% FWHM window around the signal peak. Fig.~\ref{fig:energy} shows the energy distribution of signal and background events. 

Diffusion is applied to the event hit positions. Each hit smeared according to a 3D Gaussian distribution to account for the transverse and longitudinal diffusion. The amount of diffusion applied in this sample corresponds to 0.3~$\sqrt{\text{bar}}$ mm/$\sqrt{\text{cm}}$, which is reduced from the expected diffusion from pure xenon. This choice was made to keep the file sizes manageable and produce tracks with a clarity that is more similar to deconvolved tracks in the NEXT detector as described in Ref.~\cite{NEXT2021_RLDeconvolution}. After diffusion, electrons are re-binned with a fixed voxelization of $3\times3\times3$~mm$^3$.

\begin{figure}[hbt]
\centering
\includegraphics[width=\textwidth]{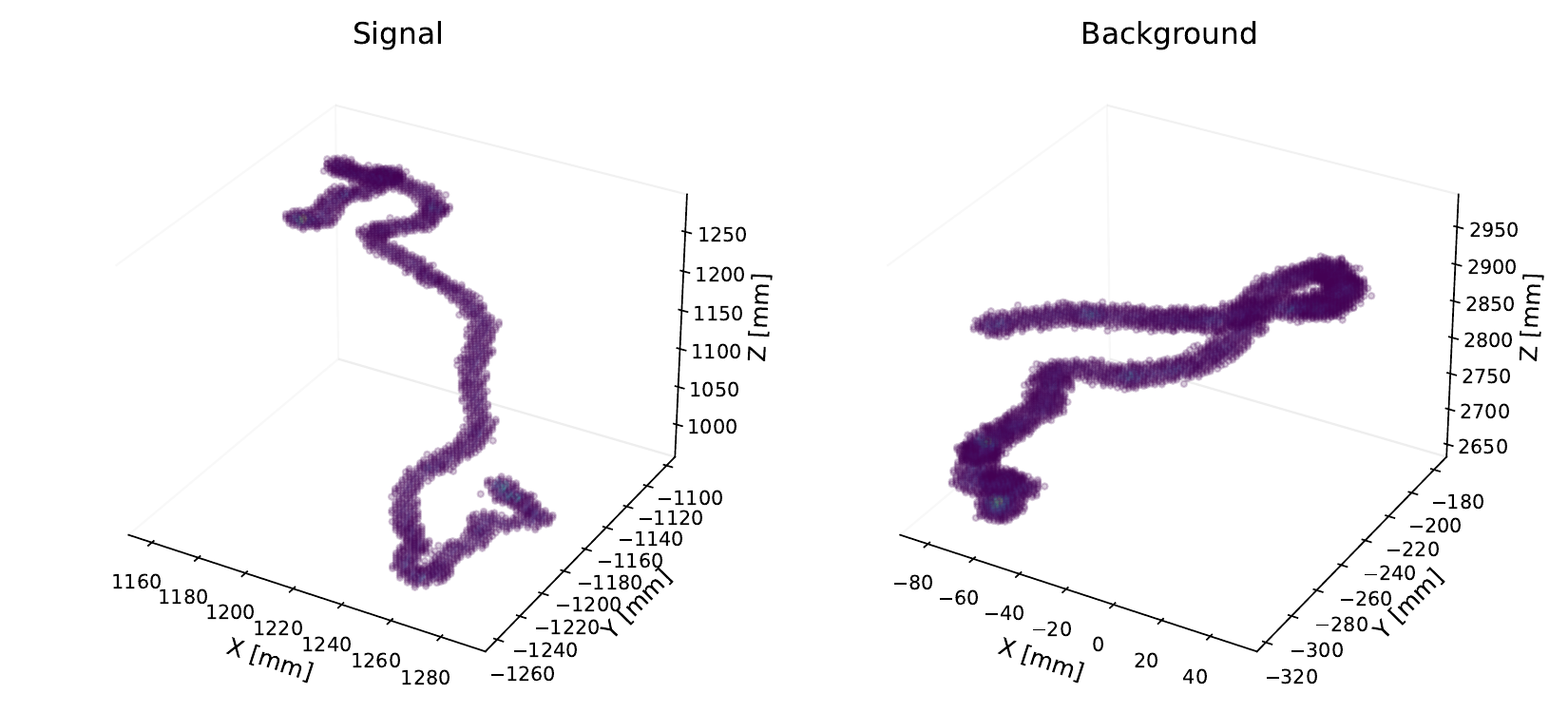}
\caption{\label{fig:event} An example event for (left) signal and (right) background event. The colour scale represents the energy deposited in a given bin. Larger energy deposits (yellow colour) can be seen towards both ends of the signal event, while the background only shows one end with a larger energy deposit.     }
\end{figure}

Example events of signal and background are shown in Fig.~\ref{fig:event}.

\section{Using the Dataset}

The dataset is stored in Zenodo at the following link:

https://doi.org/10.5281/zenodo.18927784

The dataset is stored in Zenodo in a signal (0nubb) and background (Bi) files that are split into 10 parts, which just split up the total dataset size. You can download the tar files and unzip them by doing 
\begin{lstlisting}[language=Python]
tar -xvf 0nubb_part_1.tar
\end{lstlisting}

\newpage

The files are stored in hdf5 format which can be readily loaded into a pandas dataframe:

\begin{lstlisting}[language=Python]
import pandas as pd
hits=pd.read_hdf("ATPC_0nubb_5bar_Efilt_5.0percent_smear_9.h5", "MC/hits")
print(hits)
\end{lstlisting}

Table~\ref{tab:signaldf} shows a clipped output of the dataframe for a signal event. There are six columns:
\begin{enumerate}
    \item \textbf{event\_id}: A unique integer describing each event. 
    \item \textbf{x,y,z}: the x,y,z positions of each hit in mm (a column for each). Each row is a unique hit for a given event id. 
    \item \textbf{energy}: The energy of the hit in MeV. 
    \item \textbf{label}: \textit{*Training Label*} Either Signal or Bkg to denote signal and background, respectively. 
\end{enumerate}

\begin{table}
\centering
\caption{Example layout of the signal event dataframe.}
\label{tab:signaldf}
\begin{tabular}{cccccc}
\toprule
event\_id & x [mm] & y [mm] & z [mm] & energy [MeV] & label \\
\midrule
81553 & -1249.7 & 397.3 & 815.0 & 0.000025 & Signal \\
81553 & -1243.7 & 394.3 & 810.0 & 0.000025 & Signal \\
81553 & -1241.7 & 406.3 & 805.0 & 0.000174 & Signal \\
. & . & .& . & . & . \\
. & . & .& . & . & . \\
81488 & 250.3 & -115.7 & 1443.0 & 0.000025 & Signal \\
81488 & 235.3 & -103.7 & 1446.0 & 0.000124 & Signal \\
81488 & 238.3 & -109.7 & 1446.0 & 0.000050 & Signal \\
. & . & .& . & . & . \\
. & . & .& . & . & . \\
\bottomrule
\end{tabular}
\end{table}

\subsection{Dataset Comments}

The signal size is about 8.5 GB in total, while background is 6.5 GB in total. It is not necessary to use the entire dataset for training the ML model, even 10\% of this data should be enough to get some performance. 

The 3 mm voxel size is quite small, to further reduce the dataset and increase speed in training a rebinning into larger voxels can be done. 

Plotting an event as shown in Fig.~\ref{fig:event} is simple. Simply load an event and plot using the matplotlib library:

\begin{lstlisting}[language=Python]
# Continue on from code snippet above loading the hits dataframe

import matplotlib.pyplot as plt
event_ids = hits.event_id.unique()

# You can change the index 0 to plot a different event
event = hits[hits.event_id == event_ids[0]] 

fig = plt.figure(figsize=(10,10))
ax = fig.add_subplot(axis, projection='3d')

ax.scatter(event.x, event.y, event.z, c=event.energy, cmap='viridis')
ax.set_xlabel("X [mm]")
ax.set_ylabel("Y [mm]")
ax.set_zlabel("Z [mm]")
\end{lstlisting}

\section{Disclaimer}
The NEXT Collaboration has authorized the public release of this simulation, permitting its use for all purposes. Individuals or collaborations are allowed to publish papers based on this dataset without the need to include NEXT as an author. The NEXT Collaboration maintains exclusive ownership rights over this dataset and reserves all associated rights. Should you choose to utilize this dataset in your work, NEXT kindly requests that you properly cite this Zenodo dataset.

\section{Acknowledgments}
Krishan Mistry would like to thank the NEXT collaboration for their kind permission to create and use this simulation for use at this workshop. This simulation used services provided by the OSG Consortium \cite{osg1,osg2,osg3,osg4}, which is supported by the National Science Foundation awards \#2030508 and \#1836650.

\bibliographystyle{unsrt}
{\small \bibliography{Bibliography}}
\end{document}